# Integrated Energy System Optimal Operation in Coal District With Hydrogen Heavy Trucks

*Junjie Yin*[1], *Jianhua Wang*[1]*, *Jun You*[1], *Hong Chen*[2] *and Wei Shi*[2]

[1]*School of Electrical Engineering, Southeast University, Nanjing, China,* [2]*Changzhou Power Supply Company of State Grid Jiangsu Electric Power Co., Ltd., Changzhou, China*

***Correspondence:**
*Jianhua Wang*
*wangjianhua@seu.edu.cn*

**Specialty section:**
*This article was submitted to Process and Energy Systems Engineering, a section of the journal Frontiers in Energy Research*

**Received:** *28 July 2021*
**Accepted:** *20 September 2021*
**Published:** *15 September 2021*

**Citation:**
*Yin J, Wang J, You J, Chen H and Shi W (2021) Integrated Energy System Optimal Operation in Coal District With Hydrogen Heavy Trucks. Front. Energy Res. 9:748673. doi: 10.3389/fenrg.2021.748673*

The coal industry contributes significantly to the social economy, but the emission of greenhouse gases puts huge pressure on the environment in the process of mining, transportation, and power generation. In the integrated energy system (IES), the current research about the power-to-gas (P2G) technology mainly focuses on the injection of hydrogen generated from renewable energy electrolyzed water into natural gas pipelines, which may cause hydrogen embrittlement of the pipeline and cannot be repaired. In this paper, sufficient hydrogen energy can be produced through P2G technology and coal-to-hydrogen (C2H) of coal gasification, considering the typical scenario of coal district is rich in coal and renewable energy. In order to transport the mined coal to the destination, hydrogen heavy trucks have a broad space for development, which can absorb hydrogen energy in time and avoid potentially dangerous hydrogen injection into pipelines and relatively expensive hydrogen storage. An optimized scheduling model of electric-gas IES is proposed based on second-order cone programming (SOCP). In the model proposed above, the closed industrial loop (including coal mining, hydrogen production, truck transportation of coal, and integrated energy systems) has been innovatively studied, to consume renewable energy and coordinate multi-energy. Finally, an electric-gas IES study case constructed by IEEE 30-node power system and Belgium 24-node natural gas network was used to analyze. The results show that by introducing the proposed hydrogen production technology, typical daily operating costs are effectively reduced by 7.7%. Under China's carbon emissions trading system, the operating costs of hydrogen heavy trucks have been reduced by 0.95 and 4.68% respectively, compared with electric vehicles and diesel trucks. Under Europe's stricter carbon emissions trading system, the percentages of cost reduction are 2.56 and 9.12%, respectively. The above technical results verify the feasibility, economy, low carbon, and effectiveness of the proposed mechanism.

# 1 INTRODUCTION

The coal industry is indispensable in industrial production system, and its energy consumption occupies a large share, i.e., the coal district is a typical large load of energy system. In the production process of mining, preparation, etc., coal mine machineries need to consume a lot of electricity and heat. In addition, the mined coal depends on heavy trucks to be transported to various destinations, which requires a large amount of energy to drive the engine and is essential in the coal industrial production process.

Therefore, the coal industry districts usually suffer from serious environmental pollution problems, e.g., Kuzbass (Russia), Shanxi (China), Ruhr (German), Appalachia (United States). The main reasons include but not limited to coal dust, factory emissions, vehicle exhaust, etc. Facing the tremendous environmental pressure and resource shortage, coal districts are in urgent need of energy transformation and the exploitation of renewable and indigenous resources to solve local energy use problems.

Nowadays, the development of renewable energy, represented by wind power, solar photovoltaics, etc., has become an inevitable trend. Generally speaking, coal resources are distributed in inland mountainous areas, and the local geographical features determine the abundant wind power resources in these areas. However, the phenomenon of abandoning wind power and PV is still serious, mainly due to the instability of renewable energy and the difficulty of being directly integrated into traditional energy networks.

As a clean energy, hydrogen energy is regarded as an alternative to fossil energy in the future. The current industrial hydrogen is mainly from:

1) More than 90% hydrogen is produced from fossil energy (*e.g.* , coal, natural gas) or alcohols, which emits greenhouse gases(Herraiz et al., 2020);
2) Less than 5% hydrogen is production obtained from renewable energy electrolysis, has more room for development (Pan et al., 2020). (Relatively stable wind power is more suitable than PV with greater daily volatility).

In coal districts, the significances for the development of the hydrogen are unique: volatile renewable energy can be absorbed and clean energy supply can be achieved (Wu et al., 2020). Different from the high hydrogen price in resource-deficient areas, the coal district has ample and cheap hydrogen, through the Power-to-gas (P2G) technology(Bailera et al., 2020). According to the type of conversion gas, P2G technology can be divided into two types: power-to-hydrogen (P2H) conversion and power-to-methane (P2M) conversion.

Regarding the hydrogen-based integrated energy system (IES) (Liu et al., 2021) and integrated electric-gas systems (IEGS) (Song et al., 2020; Sheng et al., 2018). The hydrogen produced in P2G plants can be used in following ways (Gahleitner, 2013):

1) Further synthesis to methane or other hydrocarbon fuels through methanation equipment;
2) Injection the mixture of hydrogen and methane into natural gas pipelines by hydrogen compressed natural gas (HCNG) technology (Zhou et al., 2020);
3) Power generation with the help of internal combustion engine or combined heat and power (CHP) devices (Ni et al., 2018);
4) Hydrogen storage into tanks after pressurization (Qiu et al., 2020);
5) Hydrogen refueling stations for vehicles or the use of hydrogen in industry.

Among the above 5 ways, way 1) does not take advantage of hydrogen energy itself, compared with other hydrocarbons. At present, some scholars are devoted to the research of way 2) – HCNG, and have achieved certain results in the study of the effect of hydrogen content percentage on integrated systems (Zhou et al., 2020). However, due to the characteristics of low density and high activity, blending hydrogen reduces the amount of energy delivered by the natural gas network under the same conditions. Hydrogen embrittlement may occur, which poses challenges to safety of the overall system. Due to the hazards of hydrogen embrittlement caused by the activity of hydrogen molecules, some studies are devoted to solving the location of hydrogen embrittlement cracks and performing subsequent repair work of steel pipelines through precise positioning (Chen et al., 2020). Therefore, countries have established strict limits on hydrogen blending in natural gas networks (generally the volume of hydrogen is up to 6%) (IEA, 2020). In terms of way 3), the direct use of hydrogen converted by P2G technology in the CHP system involves multiple energy conversions, while each conversion is bound to bring about a large energy loss, so the gain is more than the loss. As regards way 4), high cost is the primary reason hindering large-scale storage of hydrogen energy, mainly due to pressurized systems and containers. As to, the current input cost of hydrogen buses in some cities is too high to make profits, although some studies have confirmed the highlights of their proposed fuel (Bongartz et al., 2018; O'Shea et al., 2020) on light-duty vehicles(He et al., 2020).

The application of hydrogen heavy trucks is naturally proposed in this article, based on the evaluation of the above five methods. The proportion of energy consumption in transportation has remained high for a long time, and heavy trucks are an indispensable part of it. Currently, heavy trucks are mainly fueled by fossil fuels such as diesel. In recent years, hybrid electric vehicles and hybrid trolley trucks have also been applicated (Seebergs, 2014). The energy management strategies are proposed for heavy-duty fuel cell hybrid electric vehicles (Rudolf et al., 2021). For high energy-consuming mining enterprises, the system-dynamic simulation of logistics system has been studied to meet the needs of coal transportation (Mishkurov et al., 2021). Compared with the above-mentioned transportation fuel, substituting diesel with hydrogen is regarded as an effective way to reduce the air pollution and greenhouse gas emissions of heavy trucks (Lao et al., 2020). However, the above-mentioned papers do not design the industrial chain of hydrogen energy production-conversion-utilization, and the current production cost of hydrogen energy is still relatively high. Actually, conditions for the development of hydrogen heavy trucks in coal districts are superior: not only are hydrogen sources abundant, but also the application space of hydrogen heavy truck is broad. Due to the huge demand for coal transportation in these areas, the number of locally registered heavy trucks is also very large, usually reaching tens of thousands. The reasons and significance of developing hydrogen heavy trucks are as follows:

1) The rigid demand for coal transportation can be solved, skipping low-profit hydrogen fuel cell buses (Rolink and Rehtanz, 2013);
2) Fossil energy consumption can be reduced by replacing diesel heavy trucks;
3) Electrolysis using renewable energy is usually the cleanest way, which can be considered as negative greenhouse gas production;
4) Compared with lithium power batteries, hydrogen heavy trucks are more suitable for heavy-duty and long-distance transportation, and have the advantages of longer cruising range, shorter charging time, and lighter weight;
5) Hydrogen produced by electrolysis is still the main improvement direction and can get strong support from policies.

Based on the above analysis, an energy mechanism for the typical coal districts has yet to be proposed. For coal districts, there are the following visions: giving full play to the advantages of rich wind energy in coal districts, designing practical and effective energy supply structure systems, and realizing smooth, low-carbon and environment-friendly transportation of coal resources. As an energy output terminal, it has important responsibilities for the entire energy network. Therefore, the following goals should be achieved: consuming renewable energy as much as possible, breaking the status quo of the separate design and independent operation of each existing energy supply system, and establishing the IES, in order to promote the optimization of the energy structure and finally achieve the grand goal of carbon neutrality.

In order to solve the above-mentioned challenges, models of hydrogen heavy trucks and P2G equipment are established and second-order cone programming (SOCP) are applied. The main contributions of this work are summarized below.

1) The closed industrial loop between coal mining, hydrogen production, truck transportation of coal, and integrated energy systems has been innovatively proposed.
2) P2G technology has been further expanded, by taking into account the detailed chemical reaction process of P2H and P2M. Utilizing the wind power with strong volatility, the proposed P2G technology adjusts electrolysis and methanation in real time to control the generation ratio of hydrogen and methane.
3) Reasonable use of hydrogen energy, as the fuel of hydrogen heavy trucks, is proposed, which avoids hydrogen embrittlement and reduces the storage cost.
4) In the scenario of coal district, the models of hydrogen heavy truck, P2G equipment and IEGS are innovatively proposed. Incorporating heavy trucks into the unified dispatch of IEGS is conducive to making full use of hydrogen energy, reducing carbon emissions, and achieving clean and environmental protection.

The remaining parts of this paper are organized as follows. The scenario of coal district considering hydrogen heavy trucks and P2G equipment model are introduced in **Section 2**. **Section 3** describes the problem formulation and transformation. Case study is provided to prove the validity of the proposed model in **Section 4**. **Section 5** draws the conclusion on this paper.

# 2 TYPICAL SCENARIO OF COAL DISTRICT

As shown in **Figure 1**, the closed industrial loop between coal mining, hydrogen production, truck transportation of coal, and integrated energy systems has been innovatively proposed. In the coal mining process, part of the coal is used to produce hydrogen through C2H technology, and the other part is transported to power plants by hydrogen heavy trucks, which consume a lot of energy in the transportation process. Wind power is highly volatile and cannot be directly connected to the grid. Therefore, wind power produces hydrogen (energy for hydrogen heavy trucks) through electrolysis of water, and part of the hydrogen is further methanated to produce methane (injected into natural gas pipelines). Note that in this article, only the daily operating emissions (variable cost) is considered, and the emissions caused by the initial equipment investment (fixed cost) are not considered. Among them, coal is used for hydrogen production or transportation, and wind power is used for hydrogen production or methanation. By combining the needs of the coupled energy network, real-time optimized dispatching is carried out.

## 2.1 Mining, Processing and Transportation

### 2.1.1 Coal Gasification

As summarized in **Section 1**, more than 90% of hydrogen is produced by fossil energy, and the chemical reaction equation is as follows:

$$C + H_2O \xrightarrow{\triangle} CO + H_2 \quad \Delta H = +131 \text{ kJ/mol} \tag{1}$$

$$CO + H_2O \longrightarrow CO_2 + H_2 \quad \Delta H = -41 \text{ kJ/mol} \tag{2}$$

Due to the scenario of coal districts in this article, the process of producing hydrogen from fossil energy is only coal-to-hydrogen. Assuming that the weight of coal mined at the time slot $t$ is $M_t$ (ton), and $\beta_t$ (%) is the ratio of mined coal used as the raw material for coal gasification, the amount of hydrogen produced by coal in this time slot is

$$f_{H_2,t}^{Coal} = \alpha_t^{Coal}(\beta_t M_t) \tag{3}$$

where $\alpha_t^{Coal}$ represents the efficiency of coal gasification at $t$. This process is abbreviated as coal-to-hydrogen (C2H). Since the chemical reaction in (**Eq. 1**) requires heat absorption, the reaction equipment involved in the process is also one of the heat loads.

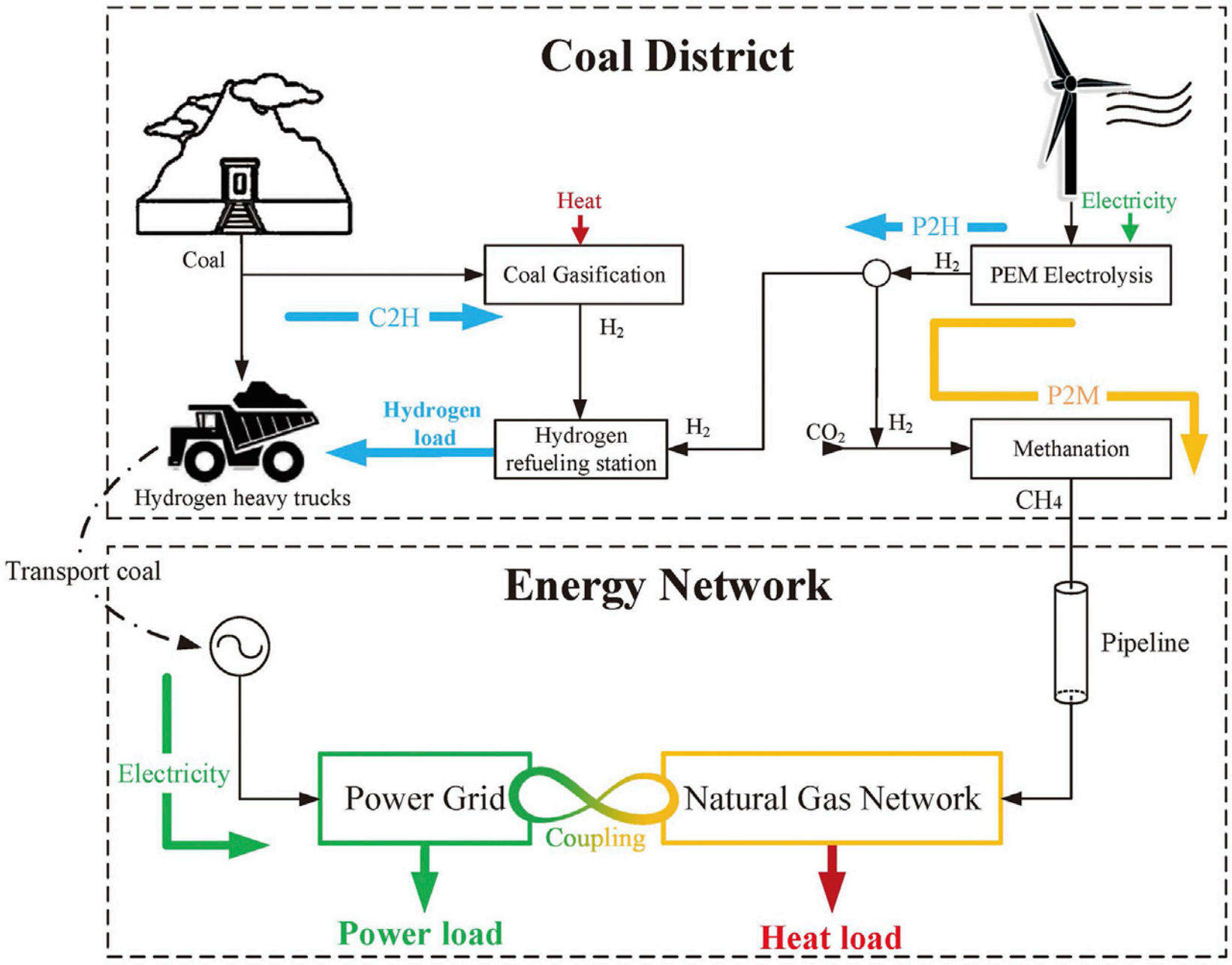

**FIGURE 1 |** Scenario of coal district considering hydrogen heavy trucks and P2G equipment model of wind power to $H_2$ and $CH_4$.

### 2.1.2 Hydrogen Heavy Trucks

With technological breakthroughs (electric stacks, membrane electrodes, catalysts, etc.), policy promotion, and the construction of hydrogen refueling stations, hydrogen fuel vehicles have a bright future. The subsidy tilt for heavy trucks is more prominent in the new policy. In the latest policy in September 2020, the upper limit of bicycle subsidies for heavy trucks over 31t has increased by 36.5% (National Energy Administration, 2020).

The sales of hydrogen heavy trucks will be almost zero before 2020, but by 2030 it is estimated that sales will account for more than 50% of the overall hydrogen-fueled vehicles. With infrastructure, technological innovations and cost reductions pushing into a period of accelerated development, sales are expected to exceed 100,000 vehicles by 2030. World-renowned automobile companies (Toyota, Hyundai, Nikola, etc.) and consulting companies (Roland Berger, etc.) have fully discussed the feasibility of hydrogen heavy trucks. In China, various provinces are also stepping up their deployment in the hydrogen energy industry. For example, Guangdong Province will build about 300 hydrogen refueling stations, and Shanghai plans to plan close to 100 hydrogen refueling stations by 2023, and more than 30 will be completed and put into operation (Roland Berger, 2021).

Currently, one of the main factors restricting the development of hydrogen heavy trucks is the cost issue: the mainstream range is 60–80 ¥/kg, and some areas require a minimum of 40 ¥/kg, and the average subsidy is 10–12 ¥/kg. At present, the energy consumption cost of hydrogen heavy trucks per 100 km is at least 400 yuan, while the energy consumption cost of the same type of diesel vehicles is 180–200 ¥/km. In the future, with the increase in scale, the average cost of hydrogen to the station is expected to be reduced to 25 ¥/kg in 2030. If the hydrogen consumption per 100 km of heavy trucks is reduced to 7 kg, the competitiveness of hydrogen fuel cell heavy trucks in terms of energy consumption will gradually begin appear. Therefore, this article focuses on the application of hydrogen heavy trucks combined with IES in coal districts to achieve cost reduction through the coordination of multi-energy flows.

**Table 1** shows the comparison between hydrogen heavy trucks and electric vehicles and diesel trucks in many aspects. It can be predicted that in the future when hydrogen energy production technology continues to increase and costs continue to fall, hydrogen heavy trucks will become more competitive.

In the actual production process, the amount of coal mined is directly proportional to the required transport capacity of heavy trucks. In this article, the load capacity and required quantity of each heavy truck are no longer considered separately. But from a global perspective, the relationship between the hydrogen consumption of hydrogen heavy trucks and the amount of coal mining is as follows:

$$f_{H_2,t}^{Truck} = \alpha_t^{Truck} (1 - \beta_t) M_t \tag{4}$$

where $f_{H_2,t}^{Truck}$ is the total hydrogen demand of heavy trucks at $t$ and $\alpha_t^{Truck}$ is the transport hydrogen consumption coefficient of trucks. $(1 - \beta_t)M_t$ represents the amount of coal that needs to be transported except for C2H.

## 2.2 P2G Equipment Model

### 2.2.1 Wind Power Electrolysis and Methanation Technology

The relationship between the wind power output $P_{wind}$ (watts) and the average wind speed $v$ (m/s) is as follows (Söder and Ackermann, 2012).

**TABLE 1 |** Comparison of heavy trucks with different power types.

| | Hydrogen truck | Electric vehicle(EV) | Diesel oil truck |
|---|---|---|---|
| Filling time | 10–15 min | Several hours | 10–15 min |
| Recharge mileage | 500–750 mile (Long-distance freight) | 100–300 mile (Short-distance freight) | 500–750 mile |
| Impact on the grid | Buffer | Relatively large | No impact |
| Energy sustainability | Promising | Depend on battery technology | Limited reserves |
| Emission | Zero | Zero | High |
| Single subsidy | $54.6 \times 10^4$ ¥ | $5.5 \times 10^4$ ¥ | $1.4 \times 10^4$ ¥ |
| Market share in 2030 | > 50% | ≈ 40% | ≈ 5% |
| Fuel weight | Very light | Relatively heavy | Heavy |
| Fuel cost in 2025 | 25 ¥/kg | 1.0–1.6 ¥/kWh | 50 ¥/t |
| Fuel consumption | 7 kg/km | 200 kWh/km | 30 L/km |

$$P_{wind} = \frac{1}{2}\rho A v^3 \tag{5}$$

where $\rho$ indicates the air density (kg/m$^3$). It can be seen from **Eq. 5** that when the wind speed is between the cut-in wind speed and the rated wind speed, the ability of the wind turbine to capture wind energy $P_{wind}$ is approximately proportional to the cube of the wind speed $v^3$.

In the current field of industrial production of hydrogen, the main technology is proton exchange membrane (PEM) electrolysers. Electric conversion to hydrogen is achieved by electrolyzing water. Emissions generated during the production of wind turbine blades are not considered in this paper. The equation of the electrolyzed water reaction is shown in **Eq. 6**.

$$2H_2O \xrightarrow{\text{electrify}} 2H_2\uparrow + O_2\uparrow \tag{6}$$

Methanation is the conversion of COx to methane $CH_4$ through hydrogenation. The reaction equations are shown in **Eq. 7** and **Eq. 8**.

$$CO + 3H_2 \xrightarrow[\text{cat.}]{\text{high temp}} CH_4 + H_2O \ \Delta H = -206 \text{ kJ/mol} \tag{7}$$

$$CO_2 + 4H_2 \xrightarrow[\text{high press}]{\text{high temp}} CH_4 + 2H_2O \ \Delta H = -164 \text{ kJ/mol} \tag{8}$$

## 2.2.2 P2H and P2M Model

P2H and P2M technology have their own characteristics:

- P2H has higher reaction efficiency than P2M;
- P2H technology conversion conditions is less difficult than P2M , and the required cost is low;
- Hydrogen cannot be injected into existing natural gas pipelines on a large scale, while methane can; • Methane is safer in view of the flammability limits (ISO, 2017) and explosion limits (ISO IEC, 2017);
- The production and combustion of hydrogen do not involve carbon emissions, and CO2 is consumed during the production of methane;
- The calorific value of hydrogen is higher, which of methane is lower.

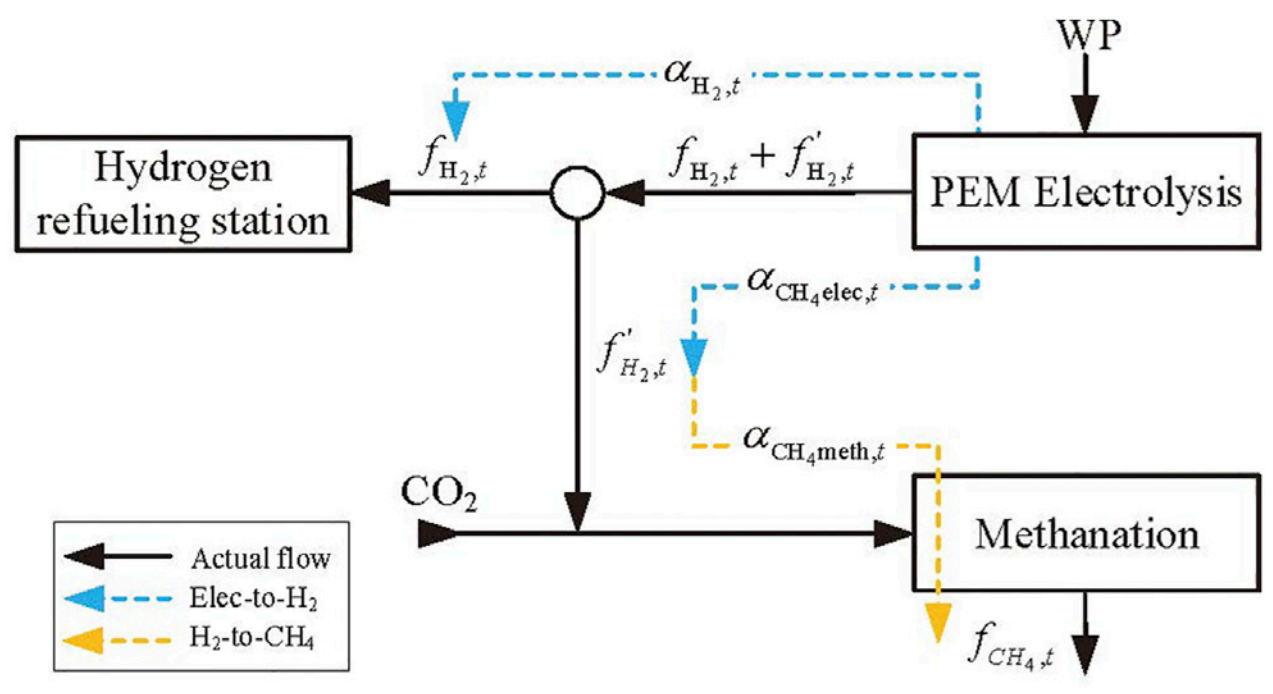

**FIGURE 2 |** Schematic diagram of actual flow and variable symbols of P2G equipment.

**1. Reaction Efficiency Constraints**

$$\begin{cases} \eta_{elec,t} > \eta_{meth,t} \\ 57\% \le \eta_{elec} \le 73\% \\ 50\% \le \eta_{meth} \le 64\% \end{cases} \tag{9}$$

where $\eta_{elec,t}$ and $\eta_{meth,t}$ are used to represent the reaction efficiency of the above two processes in the time slot $t$. Given two following assumptions:

1) In each time slot, the conversion efficiency of the electrolytic cell and methanation equipment $\eta_{elec,t}$, $\eta_{meth,t}$ remains unchanged, i.e., fixed value;
2) The chemical reaction process is a complete reaction and there is no reversible process.

**2. Conversion Power Consumption Constraint**

In the upper right corner of **Figure 1**, the actual flow of the P2G equipment is drawn separately as **Figure 2**, where each variable symbol (e.g., flow rate and energy consumption parameters) is marked.

$$Cons_{H_{2,t}} = \alpha_{H_{2,t}} f_{H_{2,t}} \tag{10}$$

$$Cons_{CH_{4,t}} = \alpha_{CH_{4elec,t}} f'_{H_{2,t}} + \alpha_{CH_{4meth,t}} f_{CH_{4,t}} = 4\alpha_{CH_{4elec,t}} f_{CH_{4,t}} + \alpha_{CH_{4meth,t}} f_{CH_{4,t}} \quad (11)$$

where $Cons_{H_{2,t}}$ represents the electric energy consumption used by the P2G equipment to produce $H_2$ for hydrogen refueling station; $f_{H_{2,t}}$ is the $H_2$ flow rate for hydrogen refueling station at the time slot $t$; $\alpha_{H_2,t}$ is the power consumption parameter corresponding to $H_2$ for hydrogen refueling station. Similarly, $Cons_{CH_{4,t}}$ is the electric energy consumed to produce $CH_4$;$f'_{H_{2,t}}$ is $H_2$ flow rate produced for further methanation; $f_{CH_{4,t}}$ is the flow rate of $CH_4$; $\alpha_{CH_{4elec,t}}$ represents the power consumption parameter of electrolysis corresponding to $CH_4$, and $\alpha_{CH_{4meth,t}}$ is that of methanation. In order to simplify the situation, only the chemical reaction described in (**Eq. 8**) is considered here, that is, carbon dioxide and 4 times of the amount of the hydrogen produce methane under the action of high temperature and high pressure and a catalyst, i.e., $f'_{H_{2,t}} = 4 f_{CH_{4,t}}$. Therefore the second equality is established in **Eq. 11**.

**3. Flammability and explosion limit constraint(ISO 10156: 2017, 2017; ISO/IEC 80079-20-1:2017, 2017)**

The minimum concentration (%) of gas in the air under flammable conditions is defined as the Lower flammability limit (LFL), and the highest is the Upper flammability limit (UFL). Similarly, Explosion limits are expressed as lower explosive level (LEL) and Upper explosive limit (UEL), respectively.

$$\begin{cases} LFL_{H_2} \le c_{H_{2,t}} \le UFL_{H_2} \\ LEL_{H_2} \le c_{H_{2,t}} \le UEL_{H_2} \end{cases} \quad (12)$$

$$\begin{cases} LFL_{CH_4} \le c_{CH_{4,t}} \le UFL_{CH_4} \\ LEL_{CH_4} \le c_{CH_{4,t}} \le UEL_{CH_4} \end{cases} \quad (13)$$

**4. Calorific Value Constraint**

The energy density of hydrogen is only about 25% of methane. When both release the same amount of energy, the volume of hydrogen is larger, and storage and transportation are more difficult. In this paper, $q_{H_2} = 119.96$ MJ/kg, $q_{CH_4} = 50.00$ MJ/kg.

# 3 PROBLEM FORMULATION

## 3.1 Minimize Long-Term Total Costs

$$\min_{\mathbf{P},\mathbf{F},\mathbf{\Pi}} C_{total} = \sum_{i=1}^{N}\left(\sum_{i=1}^{T} C_i^f(P_{i,t}) + C_i^U + C_i^D\right) \quad (14)$$

$$+ \sum_{m=1}^{N_g}\sum_{t=1}^{T} C_{m,t}^g + \sum_{i=1}^{T} C_{WP,t} + \sum_{i=1}^{T} C_{Truck,t} - \sum_{i=1}^{T} E_{Coal,t}$$

$$s.t.$$

$$(3), (4), (8) - (12)$$

$$\sum_{i=1}^{N_U} P_{i,t} = \sum_{i=1}^{N_L} P_{d,i,t} \quad (15)$$

$$\sum P_{input,i,t} = \sum P_{output,i,t} \ \forall \text{power nodes} \quad (16)$$

$$\sum_{i=1}^{N_U}\left(u_{i,t}P_{i,\max} - P_{i,t}\right) \ge \rho \sum_{i=1}^{N_L} P_{d,i,t} \quad (17)$$

$$u_{i,t} = \begin{cases} 0, \text{stop} \\ 1, \text{start} \end{cases} \quad (18)$$

$$u_{i,t}P_{i,\min} \le P_i \le u_{i,t}P_{i,\max} \quad (19)$$

$$-R_d \le P_{i,t} - P_{i,t-1} \le R_u \quad (20)$$

$$\sum_{k=t}^{t+TS-1}(1 - u_{i,t}) \ge TS(u_{i,t-1} - u_{i,t}) \quad (21)$$

$$\sum_{k=t}^{t+TO-1} u_{i,k} \ge TO(u_{i,t} - u_{i,t-1}) \quad (22)$$

$$C_{i,t}^U \ge \max\{H_i(u_{i,t} - u_{i,t-1}), 0\} \quad (23)$$

$$C_{i,t}^D \ge \max\{J_i(u_{i,t-1} - u_{i,t}), 0\} \quad (24)$$

$$P_{l,\min} \le P_{l,t} \le P_{l,\max} \quad (25)$$

$$\sum_{n|(m,n)\in A} f_{mn} = \sum_{n|(n,m)\in A} f_{nm} + s_m \quad \forall m \in \mathcal{N}_g \quad (26)$$

$$\text{sign}(f_{mn}) f_{mn}^2 \le C_{mn}^2 (p_m^2 - p_n^2) \quad \forall m,n \in \mathcal{N}_g \quad (27)$$

$$\text{sign}(f_{mn}) = \begin{cases} 1, \text{flow direction is positive} \\ -1, \text{otherwise} \end{cases} \quad (28)$$

$$\underline{s_m} \le s_m \le \overline{s_m} \quad \forall m \in \mathcal{N}_g \quad (29)$$

$$\underline{p_m} \le p_m \le \overline{p_m} \quad \forall m \in \mathcal{N}_g \quad (30)$$

where the coal consumption cost function $C_i^f$ of thermal power unit $i$ can be expressed by the following quadratic equation:

$$C_i^f(P_{i,t}) = a_i P_{i,t}^2 + b_i P_{i,t} + c_i \quad (31)$$

where $C_i^U$ represents the start-up cost of unit $i$, $C_i^D$ represents the shutdown cost of unit $i$, $C_{m,t}^g$ represents the cost of the natural gas network. For simplification, it is assumed that the punishment cost of abandoning wind power ($C_{WP,t}$) is linearly positively related to the amount of electricity abandoning. $C_{Truck,t}$ indicates the cost of truck transportation, which is positively related to the load capacity of the truck, when the load does not exceed the load capacity of the truck. $E_{Coal,t}$ indicates the profit from selling coal, which is proportional to the weight of coal transportation. Constraints **Eqs 15–25** are related to the power system, including power balance constraint, hot spare constraint, unit output constraint, unit ramp constraint, unit start and stop time constraint, start & stop cost constraint and line power flow safety constraint. And **Eqs 26–30** are related to the Natural gas system, including flow balance constraint, Weymouth equations constraint (Borraz et al., 2016), source quantity constraint and pressure range constraint.

## 3.2 Problem Transformation Based on SOCP

In the unified power flow modeling and solving, the SOCP-based power flow model has been used in network planning (Zhang et al., 2018) with binary decision variables (Zhang et al., 2021), and the problems generated are all modeled as mixed integer SOCP (MISOCP) (Badesa et al., 2020).

In order to eliminate the non-linearity caused by pressure variables, make the following variable substitutions $\pi_m = p_m^2$. Thus, the constraint (30) can berewritten as the following formula

$$\underline{\pi_m} \le \pi_m \le \overline{\pi_m} \quad \forall m \in \mathcal{N}g \tag{32}$$

The non-convexity of the steady-state flow model is derived from **Eq. 27**, where the absolute value sign $(f_{mn})$ is non-smooth and non-differentiable. To solve this problem, a pair of binary variables $f_{mn}^{+}$ and $f_{mn}^{-}$ are introduced to represent the forward and backward flow directions of the pipe $m - n$, respectively. Therefore, **Eq. 27** is equivalently replaced with

$$\left(\frac{F_{mn}}{C_{mn}}\right)^2 = \left(f_{mn}^{+} - f_{mn}^{-}\right)\left(\pi_m - \pi_n\right)\forall m, n \in \mathcal{N}g \tag{33}$$

By using bilinear relaxation, the bilinear term on the right side of (**Eq. 33**) can be replaced by (**Eqs 34–38**) (Note that since $f_{mn}^{+}$ and $f_{mn}^{-}$ are binary variables, this relaxation is strict):

$$\left(\frac{F_{mn}}{C_{mn}}\right)^2 = \lambda_{mn} \quad \forall m, n \in \mathcal{N}g \tag{34}$$

$$\lambda_{mn} \ge \pi_n - \pi_m + \left(f_{mn}^{+} - f_{mn}^{-} + 1\right)\left(\pi_m^l - \pi_n^u\right) \tag{35}$$

$$\lambda_{mn} \ge \pi_n - \pi_m + \left(f_{mn}^{+} - f_{mn}^{-} - 1\right)\left(\pi_m^u - \pi_n^l\right) \tag{36}$$

$$\lambda_{mn} \ge \pi_n - \pi_m + \left(f_{mn}^{+} - f_{mn}^{-} + 1\right)\left(\pi_m^u - \pi_n^l\right) \tag{37}$$

$$\lambda_{mn} \ge \pi_m - \pi_n + \left(f_{mn}^{+} - f_{mn}^{-} - 1\right)\left(\pi_m^l - \pi_n^u\right) \tag{38}$$

**Eq. 34** can be relaxed according to the cone format shown in **Eq. 39**, where the standard SOC formula of **Eq. 39** is expressed as

$$\left(F_{mn}/C_{mn}\right)^2 \le \lambda_{mn} \quad \forall m, n \in \mathcal{N}g \tag{39}$$

$$\left\| \begin{matrix} 2F_{mn}/C_{mn} \\ \lambda_{mn} - 1 \end{matrix} \right\|_2 \le \lambda_{mn} + 1 \tag{40}$$

# 4 CASE STUDY

## 4.1 Software and Solving Tools

In order to verify the effectiveness of the method in this paper, the above optimization program was developed using MATLAB-YALMIP platform, and the CPLEX algorithm package was used to solve the nonlinear mixed integer programming problem. The hardware environment of the test system is Intel (R) Core (TM) i7-6500M CPU @ 2.50 GHz, 8 GB RAM, Win10 64 bit (operating system), Matlab R2019b (development environment), and the YALMIP version is R2020.

IBM CPLEX ILOG is a high-performance mathematical programming solver for linear programming, mixed integer programming. The unit commitment problem is essentially a mixed integer programming (MIP) problem, which can be solved by MATLAB/CPLEX.

## 4.2 Initial Data

As shown in **Figure 3**, the data of IEEE 30-node power system is utilized, and the rest of the data is shown in **Table 2**. The IEEE 30-node test case represents a part of the American power system (located in the Midwest of the United States). The original IEEE 30-node scenario considers the processing conditions of 6 thermal power units. Here, Unit 6 connected to node 8 is changed to wind power equipment.

The natural gas network data adopts Belgium 24-node network data (Borraz et al., 2016). According to the internationally accepted natural gas sales rules, natural gas is priced according to its calorific value. The commonly used unit is Million Britain Thermal Unit (MBTU), while China's natural gas transaction is based on volume as a reference and the unit is cubic meter ($m^3$). It is known that 1 MBTU ≈ 28.3 $m^3$, and the unit conversion can be obtained: 1 \$/MBTU ≈ 0.035 \$/ $m^3$.

The maximum value of daily output power of the wind power unit is set as 15,000 kW. The fluctuation data of wind power output in each period is based on the project data of the University of Queensland in Australia. Now the unit output is increased by 30 times. Set the abandoning wind power punishment parameter $\delta_{WP} = 0.08$ \$/kWh.

## 4.3 Simulation Result

### 4.3.1 Typical Daily Operating Cost

However, because the output of a wind turbine responds to changes in wind speed relatively quickly, even within 1 min, the output of the wind turbine may change many times with changes in wind speed. Therefore, in a specified period of time, the wind turbine output calculated by using the average wind speed is equal to the average actual wind turbine output power. The output characteristics of wind turbines can be known from **Eq. 5**, the wind power used for electrolysis is proportional to the cube of the average wind speed. Wind speed is trendy and random, and the accuracy of its value plays a decisive role in the evaluation of wind energy. The wind power output data from February 27, 2019 to February 27, 2020 are selected, and the operation results in the proposed IEGS are shown in **Figure 4**. The wind power output at the same time basically conforms to winter > annual > summer, in which winter is selected from December to February of the following year, and summer is selected from June to August. Combined with the weather data measured by the wind project, wind power output is influenced by natural factors such as temperature , wind speed and so on.

On the basis of ensuring the stable operation of wind power output, in order to show the characteristics of energy-saving, clean, and fast absorption of P2G equipment, the comparison results obtained are shown in **Table 3**. Due to the independent operation of the power system and the natural gas system, and the direct connection of wind power units to the grid, the electricity-gas system lacking a coupling and mutual aid relationship cannot absorb unstable and intermittent wind power in a timely manner. In particular, the cost of abandoning wind power is penalized, leading to higher operating costs for the IEGS without P2G equipment. If the ecological benefits brought by the consumption of carbon dioxide gas by P2G equipment are taken into consideration, the advantages of P2G equipment are more prominent.

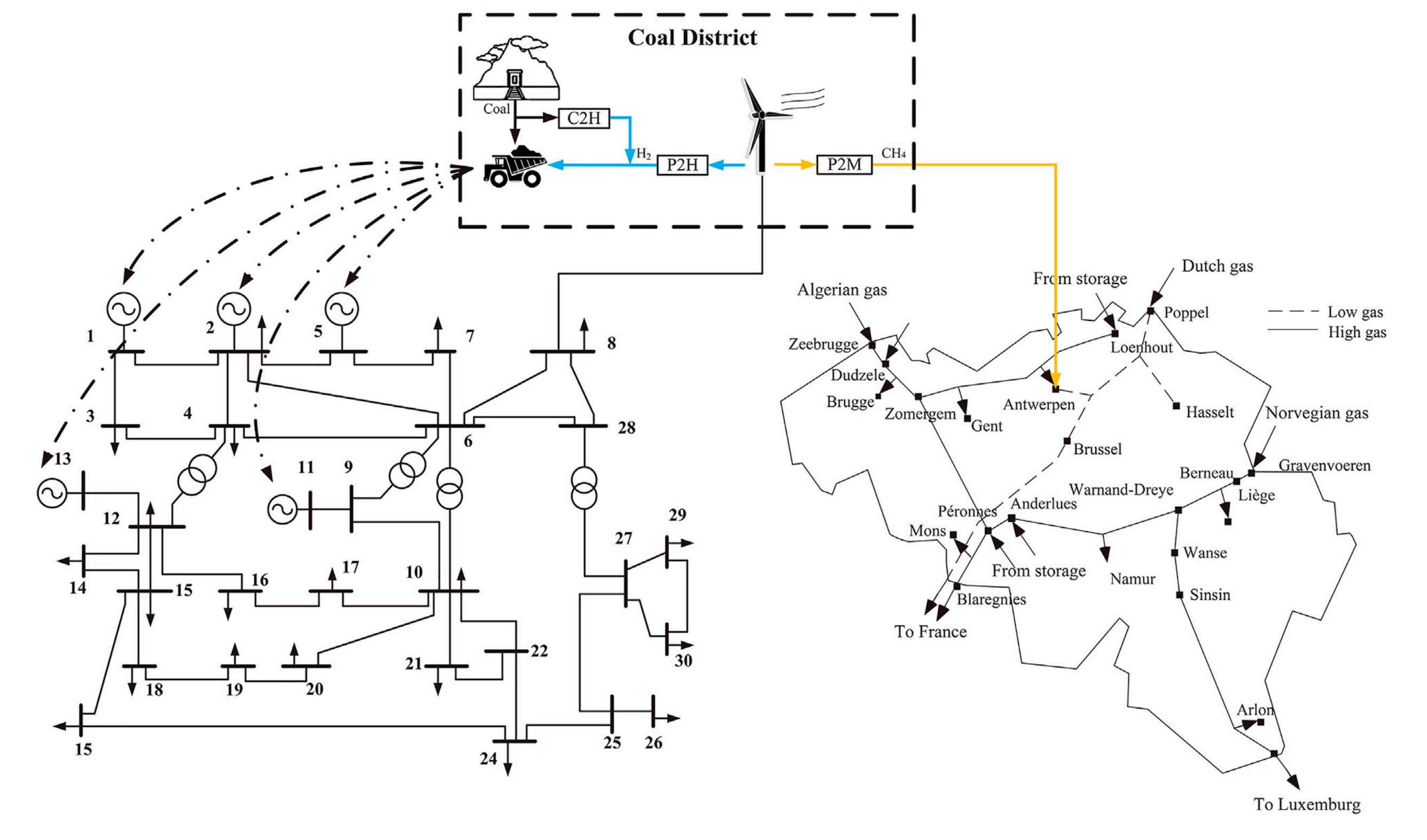

**FIGURE 3 |** IEGS case study diagram.

**TABLE 2 |** Initial data of thermal power units.

| Unit | Node | $P_{max}$ (p.u.) | $P_{min}$ (p.u.) | a (ton/(p.u.)$^2$) | b (ton/(p.u.)) | c (ton) | $R_u/R_d$ (p.u./h) | TS/TdD (h) | *H* ($/time) | *J* ($/time) |
|---|---|---|---|---|---|---|---|---|---|---|
| 1 | 1 | 1.57 | 0.50 | 01524 | 38.5390 | 786.798 | 0.37 | 2 | 3,937 | 19,686 |
| 2 | 2 | 1.00 | 0.25 | 0.1058 | 46.1591 | 945.633 | 0.30 | 2 | 25,000 | 12,500 |
| 3 | 5 | 0.60 | 0.15 | 0.0280 | 40.3965 | 1049.998 | 0.15 | 2 | 15,000 | 7,500 |
| 4 | 8 | 0.80 | 0.20 | 0.0354 | 38.3055 | 1243.531 | 0.20 | 2 | 20,000 | 10,000 |
| 5 | 11 | 0.40 | 0.10 | 0.0211 | 36.327 | 1658.570 | 0.15 | 2 | 10,000 | 5,000 |

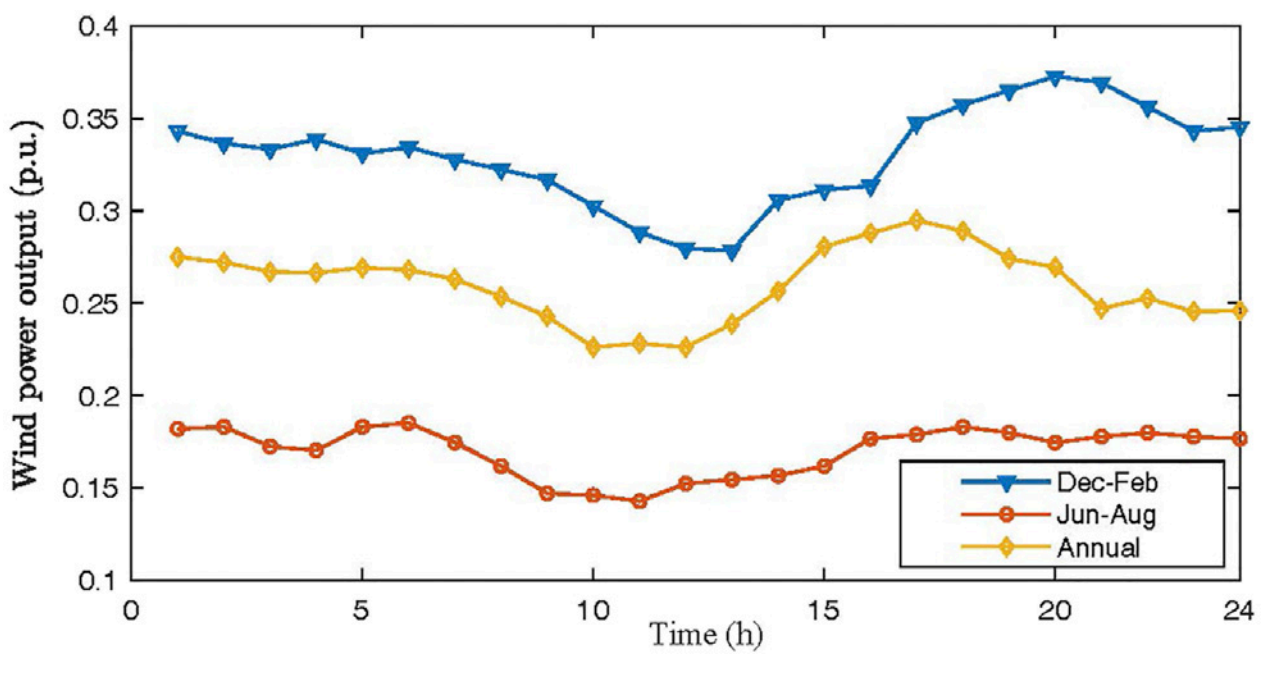

**FIGURE 4 |** Typical daily wind power output line chart.

**TABLE 3 |** Comparison of total operating costs with or without P2G.

| Total operating costs ($/day) | with P2G | without P2G |
|---|---|---|
| | $9.50763 \times 10^6$ | $10.30285 \times 10^6$ |

### 4.3.2 Multi-Energy Flow Conversion

Select 14:00 data to draw **Figure 5A**, which is based on the annual wind power output. It can be seen that among the initial energy sources, electricity and natural gas come from the power system and the natural gas system respectively, and renewable energy accounts for a considerable proportion. Through transitional energy conversion, the initial energy can be transformed into hydrogen (supplied to hydrogen heavy trucks), methane (can be directly injected into natural gas pipelines) and electricity (from power plants close to energy-consuming terminals).

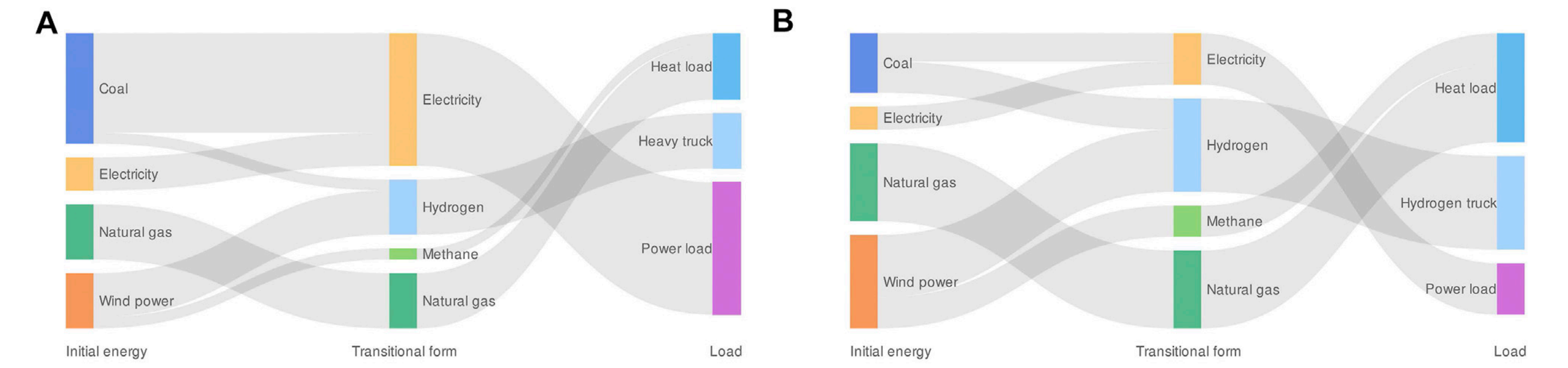

**FIGURE 5 |** The energy conversion Sankey diagram of the proposed IEGS.

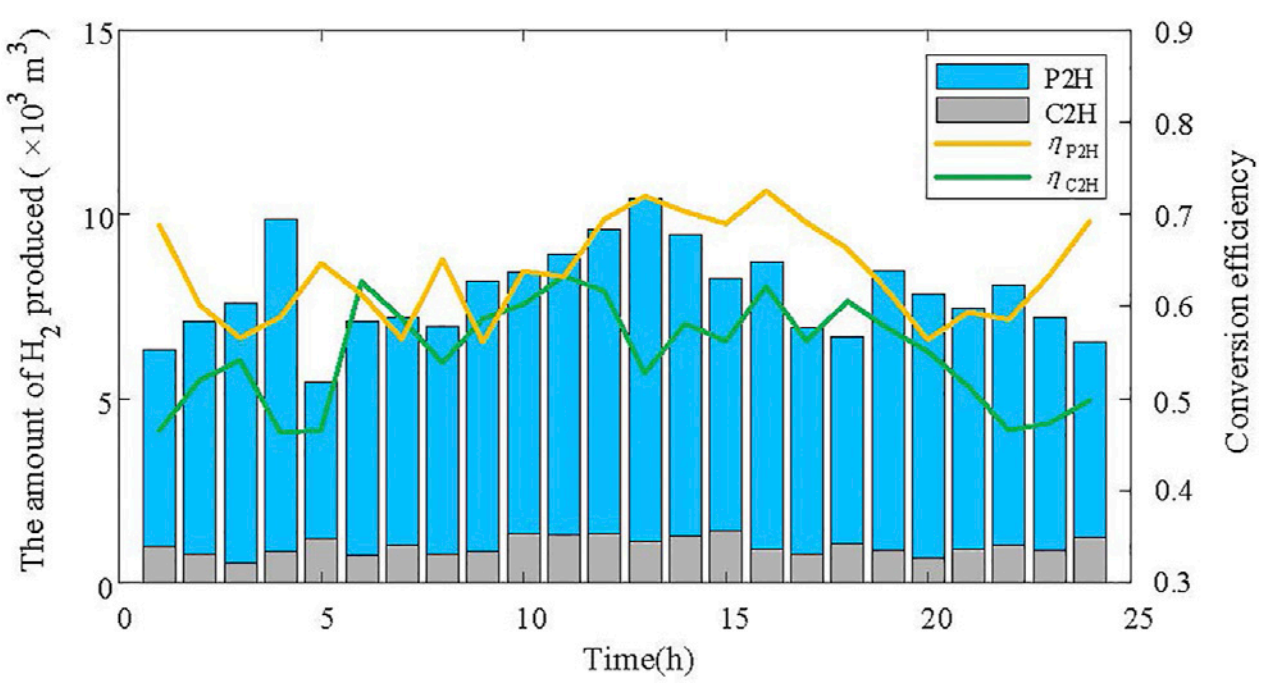

**FIGURE 6 |** Hydrogen production and conversion efficiency through P2H and C2H.

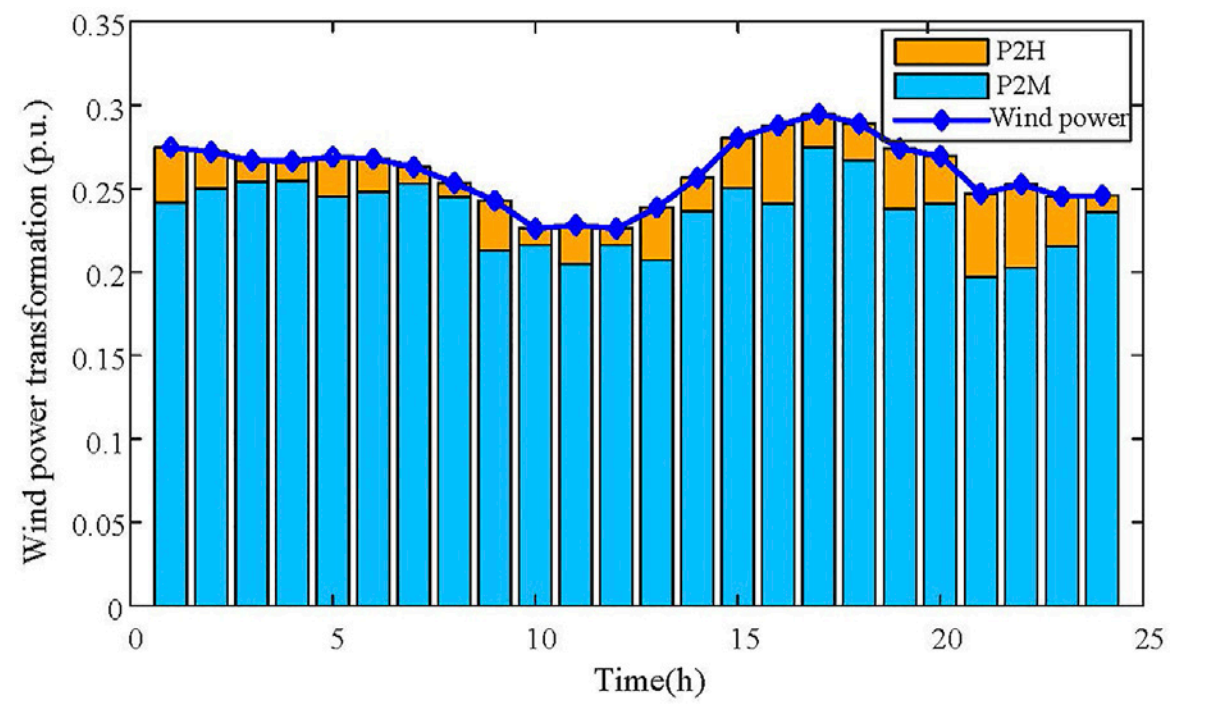

**FIGURE 7 |** Stacked bar chart of wind power converted to hydrogen and methane.

In order to compare with the above-mentioned situation, simulate the scenario with a high proportion of renewable energy installed capacity (increasing the output of the wind turbine to 2 times), and draw **Figure 5B**. With the transformation and upgrading of the energy system, coal-fired power generation is gradually eliminated due to its heavy pollution and a greater proportion of coal is used for C2H. The continuous development of technologies such as electrolysis of water to produce hydrogen has reduced the energy consumption of the conversion of the entire coal district, and the power load in this scenario has also been reduced. In view of changes in coal transportation demand, the energy produced by wind turbines meets the demand for hydrogen heavy trucks, and more hydrogen is further converted into methane, which is injected into natural gas pipelines to meet the heat load demand.

To sum up, the proposed energy utilization mechanism has certain regulatory capabilities, and can be effectively adjusted to follow changes in transportation demand, renewable energy unit output, equipment energy consumption, etc., and provide a new path for coal district transformation.

### 4.3.3 Technical Results of $H_2$ Production

Based on the hydrogen production process shown in **Figure 2**, it can be known that hydrogen production can pass through P2H and C2H. The typical daily hydrogen production and the corresponding conversion efficiency are plotted in **Figure 6**. In terms of the stacked bar chart, P2H and C2H technologies produce hydrogen, which is consumed by hydrogen heavy trucks to transport coal. The transportation volume of hydrogen heavy trucks is positively related to the required energy consumption. Because of its real-time change characteristics, the system needs to be optimized and adjusted. In detail, **Figure 5A** shows the energy conversion at 14:00, which is consistent with the corresponding value of the stacked bar chart. On the whole, the hydrogen production volume of the above two technologies is basically stable.

In terms of conversion efficiency, the conversion efficiency of P2H and C2H fluctuates relatively large due to environmental changes, price fluctuations, equipment characteristics and other factors, but they are all within a reasonable range. In contrast to the constraints of **Eq. 9**, the conversion efficiency of P2H is in the range of 57-73%. In the same way, the conversion efficiency of C2H is also within the normal working range, and there is no abnormal situation. Comparing the conversion efficiency of P2H and C2H, the former is basically higher, which is determined by complex reasons such as technical level.

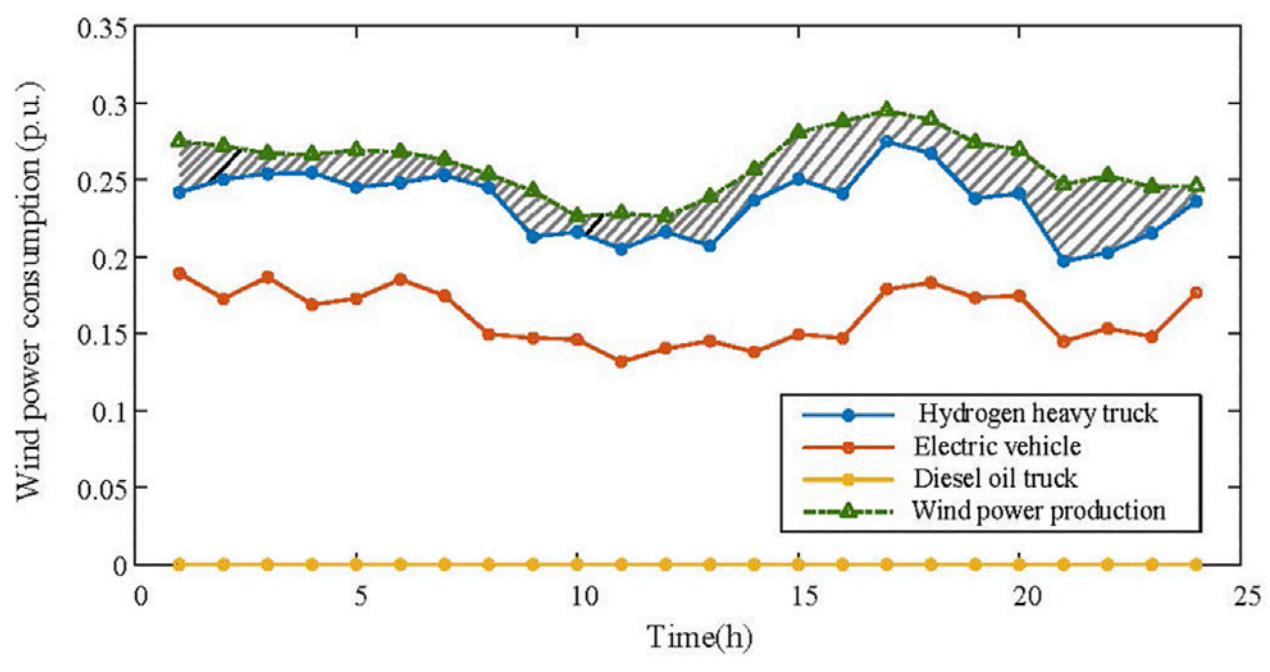

**FIGURE 8 |** Comparison of trucks consuming wind power.

Similarly, stacked bar chart of wind power converted to hydrogen and methane is drawn in **Figure 7**. The blue line in the figure represents the per-unit value of wind power output, which is derived from the annual situation in **Figure 4**. The figure shows in detail the ratio of energy generated by wind power into hydrogen and methane. The amount of hydrogen produced is related to the demand for hydrogen heavy trucks, and the remaining energy is further converted into methane. Obviously, the energy used for hydrogen production occupies a large proportion, about 5–10 times the energy used for methane production. In the future, as shown in **Figure 5B**, as the installed capacity of renewable energy increases, a greater proportion of wind power will be used for methane production.

### 4.3.4 Technical Results of Trucks Comparison

**Figure 8** shows the consumption of wind power by heavy trucks driven by different fuels. In the figure, the green dotted line represents the output of wind power. Using the mechanism proposed in the article, the hydrogen heavy truck (blue line) can basically absorb a large amount of wind power in time. Electric vehicles (red lines) can only absorb part of the renewable energy, which may involve complex conversion and storage of electrical energy with large delays. However, Diesel oil truck (yellow line) does not involve the conversion of renewable energy and cannot absorb any wind power by itself. It is worth noting that the gray shaded part between the green line and the blue line in the figure is the stable difference between the wind power output and the absorbed energy of the hydrogen heavy truck, which is exactly the energy consumption of P2M in **Figure 7**.

In order to alleviate the greater environmental pressure and implement carbon neutral initiatives, carbon emissions trading is considered in this section (Ricke et al., 2018; Zhang et al., 2020). According to the Status Report 2019 of International Carbon Action Partnership (ICAP), emissions trading systems of most countries in the world currently are in force, scheduled or under consideration. This article uses the carbon emissions trading of China and the European Union as a measurement standard, and calculates the operating costs of the three heavy trucks mentioned in **Section 2.1.2**. Assuming that emissions from renewable energy electrolysis at the production stage is taken out of study scope.

According to the results of carbon emissions trading in 2020, prices in China are about 30–40 ¥/t$CO_2$, but that in the EU has exceeded 30 €/t$CO_2$. Taking into account carbon emissions trading, the revised total cost is $(C_{total} - C_{CO_2})$, where $C_{CO_2}$ indicates the cost of $CO_2$ emissions and proportional to the amount of $CO_2$ emissions. According to **Table 4**, under China's carbon emissions trading system, the operating cost of hydrogen heavy trucks compared to electric vehicles is reduced by 0.95%, which is 4.68% compared to diesel oil trucks. Under Europe's stricter carbon emissions trading system, the percentages of cost reduction are 2.56 and 9.12%, respectively. Horizontally comparing the operating costs of hydrogen heavy trucks in China and the European Union, the operating costs in China are slightly higher by 1.31%. The result shows that diesel oil trucks or LNG trucks will be levied high carbon emission fees, which will continue to increase as carbon neutrality advances. The current price of hydrogen heavy trucks is basically similar to that of electric vehicles, because in the proposed scenario, the methanation process can absorb a part of $CO_2$. Therefore, hydrogen heavy trucks can not only achieve zero emissions on the basis of electric vehicles, but also absorb greenhouse gases, which are converted into profits here. With the continuous breakthrough of hydrogen energy technology and the inclination of government policies, the price of hydrogen energy will gradually drop and the scale of production will gradually expand. Further, hydrogen heavy trucks will have a broader prospect.

**TABLE 4 |** Comparison of total operating costs considering carbon emissions trading.

| Total operating costs ($/day) | China | European Union |
|---|---|---|
| Hydrogen truck | $9.47361 \times 10^6$ | $9.34901 \times 10^6$ |
| Electric vehicle | $9.56402 \times 10^6$ | $9.58834 \times 10^6$ |
| Diesel oil truck | $9.91750 \times 10^6$ | $10.20187 \times 10^6$ |

**TABLE 5 |** Comparison of total operating costs with different abandoning wind power punishment parameters.

| Abandoning wind power punishment parameters ($/kWh) | Total operating costs ($/day) |
|---|---|
| 0.01 | $10.02773 \times 10^6$ |
| 0.05 | $9.54392 \times 10^6$ |
| 0.08 | $9.50763 \times 10^6$ |
| 0.12 | $9.84057 \times 10^6$ |

### 4.3.5 Unit Output Curve

When the hot spare coefficient $\rho$ is 0.05, the cumulative output ladder diagram of the unit is shown on the left side of **Figure 9A**, which is basically in line with the distribution characteristics of power consumption peaks and valleys in reality, that is, there is a peak power consumption around noon during the day. In order to characterize the output status of each unit, draw the ladder diagram on the right side of **Figure 9A**. Each curve is basically stable, and there are no several start and stop conditions within a short period of time, which plays a good role in the overall stability of the system.

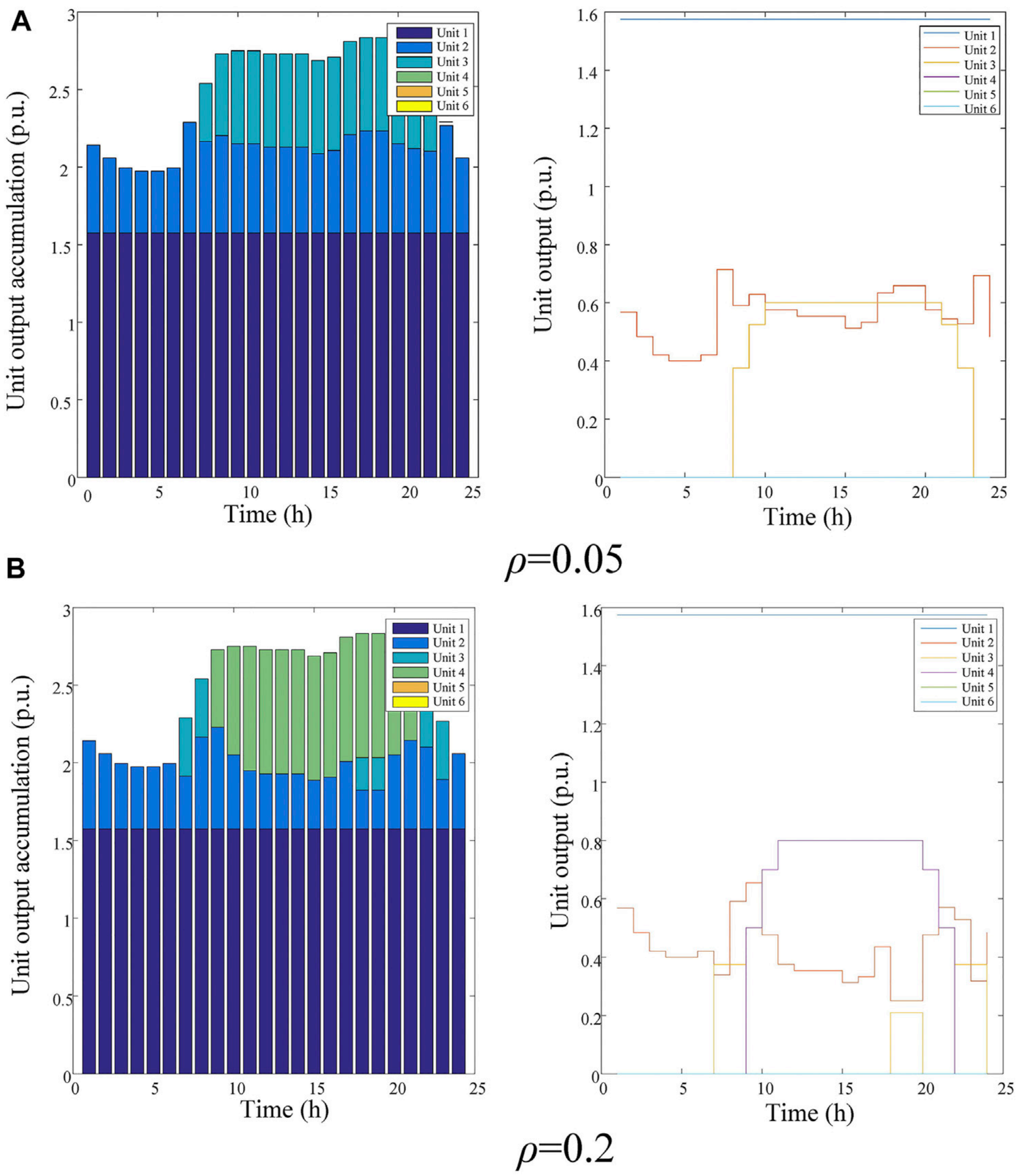

**FIGURE 9 |** Unit output curves under different hot spare coefficient.

Similarly, when the unit hot spare coefficient is 0.2, **Figure 9B** are available. The cumulative output ladder diagram of the unit is shown on the left side of **Figure 9B**. The ladder diagram on the right side of **Figure 9B** shows the output status of each unit and the curves are basically stable. However, some problems can be found in comparison with $\rho$=0.05. The right side of **Figure 9B** shows that the output of one unit has a short start and stop near 19:00, which is caused by the higher hot spare. Under the premise that a certain unit operating life and system stability are lost, the system's ability to respond to emergencies can be greatly improved.

### 4.3.6 Sensitivity Analysis of Abandoning Wind Power Punishment Parameters

In this section, adjust the abandoning wind power punishment parameters appropriately, set to four values of 0.01, 0.05, 0.08, and 0.12 respectively. The operation results of IEGS are shown in **Table 3**. When the abandonment punishment parameter is too small, the operating cost is higher, second only to the operating cost under the condition of no P2G equipment, indicating that even if the punishment parameter is small, the amount of abandonment is large, which still leads to the system. The overall operating cost is high. With the increase of the abandonment punishment parameters, the operating cost of the system has been significantly reduced, which proves that considering the abandonment punishment cost in the optimization objective can make the system fully absorb wind power. However, if the punishment parameters for abandoning solar energy continue to increase, the operating cost of the system will increase. This is because a small amount of unconsumed wind power will be multiplied by a larger punishment coefficient, which will cause a certain punishment cost.

Further, as shown in the dotted line in **Figure 10**, it can be estimated that although there is a nonlinear relationship between the abandoning wind power punishment parameter and the operating cost, there should be a minimum value as shown in **Figure 10**. When the abandonment punishment parameter takes this value, the total operating cost of the system is the smallest, and the specific value needs to be further explored. Moreover, when the abandoning wind power punishment parameter is less than the optimal value, the curve drops faster, and the difference between the values of the abandoning wind power punishment parameter is not very large, but a large amount of abandonment energy multiplied by it will cause a larger punishment cost. Correspondingly, when the abandoning wind power punishment parameter is greater than the optimal value, even if the abandoning wind power punishment parameter becomes significantly larger, the cost curve rises slowly because the amount of abandonment remains at a small amount.

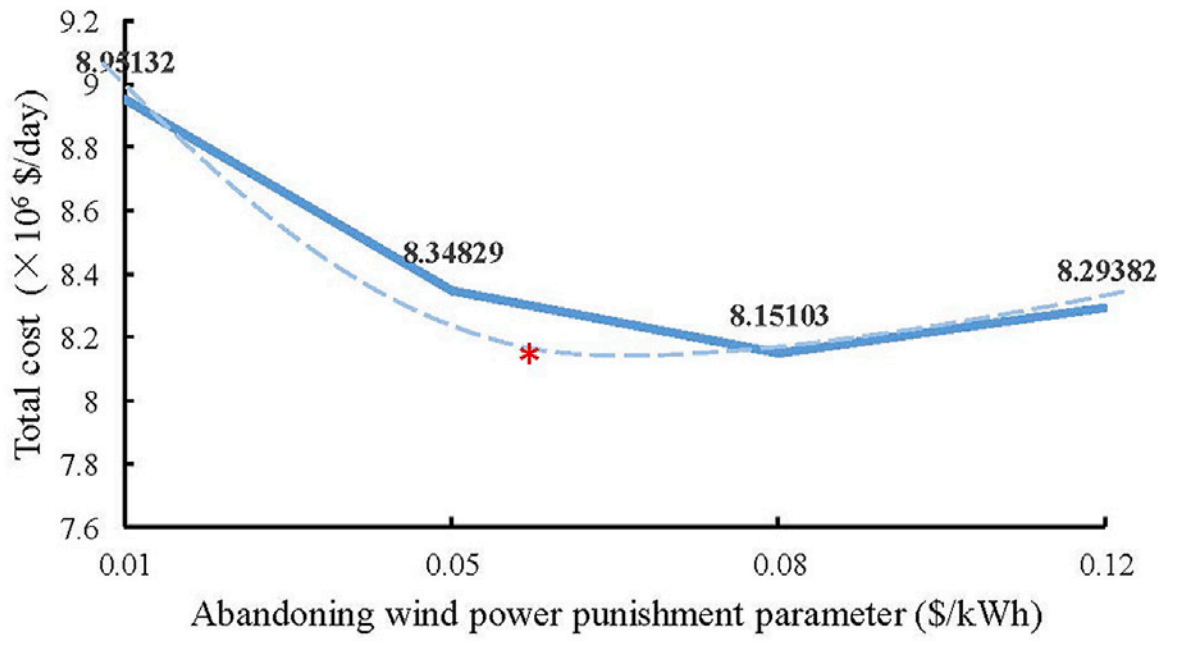

**FIGURE 10** | Curve line chart of the influence of $\delta_{WP}$ on $C_{total}$.

## 5 CONCLUSION

In view of the gradual transformation of the energy structure and the continuous development of the Energy Internet, IESs are important research direction. This paper draws the following conclusions:

1) Typical daily operating cost is significantly reduced by 7.7%, by introducing the P2G model which converts seasonally volatile wind energy into hydrogen and methane.
2) Hydrogen heavy trucks have certain advantages in the proposed coal districts scenario, *e.g.*, zero emissions and indirect consumption of renewable energy. Especially after the introduction of carbon emissions trading, the cost advantage is more prominent, compared with electric vehicles and diesel oil trucks.
3) The proposed P2G equipment and C2H technology improve the ability to absorb wind power, enhance the reliability of the system, and solve the problem of volatility. More importantly, the mechanism in the article circumvents the potential risks caused by hydrogen mixing in natural gas pipelines, the cost of which cannot be measured.
4) The abandoning wind power punishment parameter is considered in this article, and the optimal parameter is formulated according to the actual situation, which plays a certain guiding role for the future energy market leverage.

## DATA AVAILABILITY STATEMENT

The original contributions presented in the study are included in the article/Supplementary Material, further inquiries can be directed to the corresponding author.

## AUTHOR CONTRIBUTIONS

JYi: Conceptualization of this study, Methodology, Software, Formal analysis, Investigation, Data Curation, Writing - Original Draft,Visualization. JW: Validation, Resources, Writing - Review & Editing, Supervision, Project administration. JYo: Writing - Review & Editing, Visualization, Supervision, Project administration. HC: Writing - Review & Editing, Resources. WS: Writing - Review & Editing, Resources.

## FUNDING

This work was supported in part by the Science and Technology Program of State Grid Jiangsu Electric Power Co., Ltd. (J2019082) and Natural Science Foundation of Jiangsu Province (BK20181283).

**Conflict of Interest:** Author HC and WS are employed by Changzhou Power Supply Company of State Grid Jiangsu Electric Power Co., Ltd.

The remaining authors declare that the research was conducted in the absence of any commercial or financial relationships that could be construed as a potential conflict of interest.

## NOMENCLATURE

**C2H** Coal-to-hydrogen of coal gasification

**IEGS** Integrated electric-gas system

**IES** Integrated energy system

**P2G** Power-to-gas

**P2H** Power-to-hydrogen

**P2M** Power-to-methane

**SOCP** Second-order cone programming Parameters and constants

$\delta WP$ Abandoning wind power punishment parameter

$\rho$ Hot spare coefficient

**LEL** Lower explosive limit of $H_2/CH_4$

**LFL** Lower flammability limit of $H_2/CH_4$

**UEL** Upper explosive limit of $H_2/CH_4$

**UFL** Upper flammability limit of $H_2/CH_4$

$C_{i,t}^{U/D}$ Start/Stop cost of unit $i$ at time $t$

$C_{mn}$ Weymouth constant

$H_i$, $J_i$ Single start / stop cost of unit $i$

$R_d$, $R_u$ Unit down / up ramping speed

**TS** TOMinimum stop / start time Sets and indices

$\mathcal{N}_{\mathcal{G}}$ Set of natural gas nodes, $\mathcal{N}_{\mathcal{G}}$ = {1, 2, ..., m, ..., Ng}

$\mathcal{N}_{\mathcal{L}}$ Set of load nodes, $\mathcal{N}_{\mathcal{L}}$ = {1, 2, ..., i, ..., NL}

$\mathcal{N}_{\mathcal{U}}$ Set of thermal units, $\mathcal{N}_{\mathcal{U}}$ = {1, 2, ..., i, ..., NU} Variables

$\alpha_t^{Truck}$ Transport consumption coefficient of hydrogen trucks

$\rho$ Hot spare coefficientAir density

$f_{H2,t}^{Coal}$ Flow rate produced by C2H at $t$

$f_{H2,t}^{Truck}$ Total hydrogen demand of heavy trucks at $t$

$P_{wind}$ Wind power output

$v$ Average wind speed

$\alpha_t^{Coal}$ Efficiency of coal gasification at $t$

$\alpha_{CH_{4elec,t}}$ Power consumption parameter of electrolysis corresponding to $CH_4$

$\alpha_{CH_{4meth,t}}$ Power consumption parameter in process of methanation corresponding to $CH_4$

$\alpha H_2, t$ Power consumption parameter

$\beta_t$ Ratio of mined coal used as the raw material for coal gasification at $t$

$\eta_{elec,t}$ Reaction efficiency of electrolysis at $t$

$\eta_{meth,t}$ Reaction efficiency of methanation at $t$

$c_{H_2/CH_4}$ Concentration percentage of $H_2/CH_4$

$\pi_m$ Square of the pressure value at node $m$

$\underline{\pi_m}$, $\overline{\pi_m}$ Threshold of the square of the pressure value of $m$

$\underline{p_m}$, $\overline{p_m}$ Threshold of pipeline pressure

$\underline{s_m}$, $\overline{s_m}$ Threshold of natural gas source storage capacity

$C_{CO_2}$ Cost of $CO_2$ emissions

$C_i^f$ Coal consumption cost of thermal power unit $i$

$C_{m,t}^g$ Cost of the natural gas network

$C_{WP,t}$ Punishment cost of abandoning wind power

$C_{Truck,t}$ Cost of truck transportation

$Cons_{H_2,t}$ Electricty consumption of producing $H_2$

$E_{Coal,t}$ Profit from selling coal

$f'_{H_2,t}$ $H_2$ flow rate produced by P2G equipment for further methanation at $t$

$f_{H_2,t}$ $H_2$flow rate produced by P2G equipment for hydrogen refueling station at $t$

$f_{mn}$ Natural gas flow from node $m$ to node $n$

$M_t$ Weight of coal mined at $t$

$p_m$ Pressure value of node $m$

$P_{d,i,t}$ Load demand for electricity of node $j$ at time $t$

$P_{i,\max}$ Maximum output value of unit $i$

$P_{i,\min}$ Minimum output value of unit $i$

$P_{i,t}$ Output of thermal power unit $i$ at time $t$

$P_{input/output,i,t}$ Power flow input/output of node $i$ at $t$

$P_{l,t}$ Power flow of line $l$ at time $t$

$s_m$ Gas flow directly injected into the node $m$ from the source

$u_{i,t}$ Start and stop status of unit $i$ at time $t$